\documentclass[doublecol]{epl2}

\title{Friedel  oscillations of  screening  in  nanotubes}
\shorttitle{Friedel  oscillations of  screening  in  nanotubes} 

\author{A.V. Chaplik\inst{1,2} \and L.I. Magarill\inst{1,2} \and R.Z. Vitlina\inst{1}}
\shortauthor{A.V. Chaplik \etal}

\institute{
  \inst{1} Institute of Semiconductor Physics SB RAS,
Novosibirsk, 630090, Russia\\
  \inst{2} Novosibirsk State University,
Novosibirsk, 630090, Russia }

 \pacs{73.63.Fg}{Nanotubes}

\abstract{ In 3D and 2D electronic systems the singular
contribution to the static permittivity $\epsilon$  (Kohn
singularity) is a small correction to the regular part of
$\epsilon$ but it results in the leading term in asymptotic
behavior  of the screened potential (Friedel oscillations). In the
present letter we show that for nanotubes quite different results
are valid: $\epsilon$ becomes infinitely large at the singular
point and the Friedel oscillations do not play the dominant role
in the screening at the large distances. Moreover, the zero and
highest cylindrical harmonics of the effective potential are
screened by quite different mechanisms.}

\begin{document}

\maketitle

 Collective effects in quasi-1D systems are remarkable by a
number of distinguishing features. Probably the most non-trivial
one relates to the strictly 1D system of interacting fermions with
the linear single particle dispersion law (the Luttinger model).
However, there exist also different 1D objects not described in
terms of the Luttinger liquid first of all due to necessity to
account for the transversal subbands and transitions between them.
The corresponding example is a nanotube - a hollow cylinder with
2D electron gas on its surface. As usually, availability of the
mobile electrons results in renormalization of the electron -
electron interaction, i.e., screening. The specific form of the
screened e-e interaction is determined by the effective dimension
and by the energy spectrum of the electron gas. As is known in 3D
plasma the bare Coulomb interaction is replaced (due to screening)
by the Yukawa law plus the Friedel oscillations
$\cos{(2p_Fr)}/r^3$ if the electron gas is degenerate; here $p_F$
is the Fermi momentum, $r$ is the distance from a point charge,
and $\hbar=1$. Similar results in the 2D case read: the regular
part of the screened potential at $r\rightarrow\infty$ is
$Qa_B^2/r^3$ ~\cite{ando}, where $Q$ is the initial point charge,
$a_B$ is the effective Bohr radius, whereas the oscillating
contribution is proportional to $\cos(2p_Fr-\pi/4)/r^2$. We see
that both in 3D and in 2D problems the Kohn singularity of
$\epsilon(k)$ at $k=2p_F$ gives the leading term of the screened
potential for $r\rightarrow\infty$; here $\epsilon(k)$ is the
static permittivity depending on the momentum $k$.

        Consider now a nanotube with the semiconductor type of the single particle energy spectrum:
 \begin{equation}\label{en_spec}
   \varepsilon_{p,l}=\frac{p^2}{2m} +Bl^2; ~~  B\equiv\frac{1}{2ma^2},~~~~ l=0,\pm1,\pm2.....
    \end{equation}
Here $p$ is the electron momentum parallel to the tube axis, $l$
is the number of the subband, $m$  is the effective mass and $a$
is the nanotube radius. Thus, the problem in question  differs
from the similar one for a planar 2D system by only quantization
of one of the component of momentum: $p_x=l/a$.
    The considered system is infinite and uniform in $z$-direction
    (the nanotube axis)  and periodic and  uniform in the azimuthal
    direction  $\varphi$ . Hence, the electron Green function $G$     depends
    on only differences $z-z',\varphi-\varphi'$. In
    the Fourier representation $G$     is
    diagonal: $G(p,p';l,l')=\delta_{pp'}\delta_{ll'}G(p,l)$.

In what follows we will apply the standard diagram technic to find
the $e-e$ interaction renormalized by screening and we will use
the {\it linear} theory of screening. We are aware that in
strictly 1D system with linear single-particle dispersion law the
Luttinger liquid model is valid. The linear screening theory
relates to only long-wave length limit. This theory breaks down
for the momentum of the order of $p_F$ just when the Friedel
oscillations become essential
~\cite{egger,yuan,grab}.\footnote{There is a misprint in the
eq.(8) of  ref. \cite{egger}. Asymptotic
 behavior of the screened potential $U_{eff}$ must be
 $1/|x|\ln^2|x/d|$ rather than $1/|x|\ln|x/d|$, otherwise
 $\int U_{eff}(x)dx$ diverges while the Fourier component $\overline{U}_{eff}(q)$
 remains finite at $q\rightarrow 0$.}

However we actually deal with 2D object and we use the parabolic
dispersion law for electrons. For such a situation the Luttinger
 model is not applicable and we do not have any other instrument
 except RPA to investigate the problem. After that it is not
 surprising that our results for the Friedel contribution to the
 screened potential differ qualitatively from the ones obtained in
 ~\cite{egger,yuan,grab} and also ~\cite{schulz}.

 The main difference is dependence of our effective (screened)
 potential not only on $z$ (coordinate along the tube axis) but
 also on $\varphi$ - azimuthal separation between two electrons on
 the surface of a hollow cylinder. We discovered that average
 potential (zero cylindrical harmonic) and all the other harmonics
 are screened qualitatively different.

  We guess that in the present situation - there is an exactly
  solvable model but for only strictly 1D system with linear
  dispersion law and an approximate method for general case - it
  is worth to find general results within RPA and to see what
  experiments will show. This will be done in what follows.

 The Gell-Mann-Brueckner \cite{gell} chain of electron loops
  determining the Fourier component of the screened e-e interaction
  $V(k,n)$ comes to the geometrical progression:
\begin{equation}\label{V}
    V(k,n)=\frac{V^{(0)}(k,n)}{1+V^{(0)}(k,n)\Pi(\omega;k,n)},
    \end{equation}
where $V^{(0)}$ is the bare Coulomb interaction:
 \begin{eqnarray}\label{V0}
    V^{(0)}(k,n)=\tilde{e}^2\int_{-\infty}^\infty\int_0^{2\pi}\frac{e^{-ikz-in\varphi}dz d\varphi
    }{\sqrt{z^2+4a^2\sin^2{(\varphi/2)}}} = \nonumber\\ =4\pi\tilde{e}^2I_n(|k|a)K_n(|k|a),
\end{eqnarray}
Here $\tilde{e}^2=e^2/\chi$, $\chi$ is the background dielectric
constant,  $I_n, K_n$  are the modified Bessel functions of the
1-th and the 3-rd type. The polarization operator (electron loop)
has a form
 \begin{eqnarray}\label{Pi}
    \Pi(\omega;k,n)
    =\frac{1}{2\pi^2 }\sum_{l=-\infty}^\infty\int_{-\infty}^\infty dp\frac{f_{p-k,l-n}-f_{p,l}}{\varepsilon_{p,l}-
    \varepsilon_{p-k,l-n}-\omega-i\delta} \\
     (\delta=+0), \nonumber
    \end{eqnarray}
where $f_{p,l} \equiv f(\varepsilon_{p,l})$ are the Fermi
occupation  numbers.

    We are interesting here in the static
screening and we put from now on $\omega=0$. By transforming
$V(k,n)$ from eq.~(\ref{V}) back to $z$-space we obtain the
expansion of the screened interaction in cylindrical harmonics. As
$V^{(0)}(k,n)$ and $\Pi(0;k,n)$ are even functions of $n$ we come
to the series
\begin{eqnarray}\label{V(z,phi)}
    V(z,\varphi)=\int_{-\infty }^{\infty }
    \frac{dk}{(2\pi)^2}\exp{(ikz)}
    (V(k,0)+\nonumber \\
    +2\sum_{n=1}^{\infty}V(k,n)\cos{(n\varphi)}).
    \end{eqnarray}

The slowest decreasing term at $|z|\rightarrow\infty$ is given by
the zero harmonic and reads
\begin{eqnarray}\label{V0(z,phi)}
    V_0(z)=
    \int_{0}^\infty \frac{dk}{\pi}\cos{(kz)}\frac{4\pi \tilde{e}^2 I_0(ka)
    K_0(ka)}{1+4\pi \tilde{e}^2 \Pi_0(k)I_0(ka)K_0(ka)},
        \end{eqnarray}
where
    \begin{equation}\label{Pi0}
    \Pi_0(k)\equiv\Pi(0;k,0)=\frac{m}{\pi^2
    k}\sum_{-L}^{L}\ln{\biggl|\frac{k+2p_l}{k-2p_l}\biggr|},
    \end{equation}
$L$ is the number of the highest occupied subband at zero
temperature, $p_l=\sqrt{2m(E_F-Bl^2)}$ ~~~ is the Fermi momentum
of the $l$-th subband. \ \ Eq.~(\ref{Pi0}) ~~~ demonstrates the
qualitative difference of the quasi-1D problem and the 2D and 3D
problems. In 2D and 3D systems the Kohn singularity at $k=2p_F$
gives small corrections to the regular part of the polarization
operator: $\xi\ln{|\xi|}$ in 3D \cite{lifshitz} and $\sqrt{|\xi|}$
in 2D case, where $\xi=(k-2p_F)/2p_F\ll 1$. But in 1D situation
$\Pi_0(k)$ becomes infinitely large at $k=2p_l$ and the
renormalized interaction vanishes at these points.

    To calculate $V_0(z)$ in the regime $|z|>>a$ we write in the eq.~(\ref{V0(z,phi)})
$\cos{(kz)}=\mbox{Re}(\exp(ikz))$, and turn the path of
integration to the upper imaginary semiaxis. At large $|z|$
     we have two contributions to  $V_0(z)$: $\overline{V}_0$ from the point
     $k=0$,
       where $K_0$ has logarithmic singularity but $\Pi_0$  remains finite, and a number of the
        Kohn contributions from the points $k=2p_l$ resulting in the Friedel
        oscillations  $\widetilde{V}_0(z)$.
 For the non-oscillating part $\overline{V}_0(z)$ we get the expansion in inverse powers of the
  value $\Lambda\equiv\ln{(2|z|/a)}-C$          (  $C$  is the Euler constant)

  \begin{equation}\label{barV}
    \overline{V}_0(z)=\frac{\tilde{e}^2}{z}\bigl (\frac{ma_B}{4\pi\kappa_0 \Lambda}\bigr)^2 \biggl(1-\frac{ma_B/\pi \kappa_0
     + 4C}{2\Lambda}+\cdots \biggr),
    \end{equation}
    where $\kappa_0\equiv\Pi_0(k\rightarrow 0)=m[1/p_0+
    2\sum_{l=1}^L(1/p_l)]/\pi^2$; $\pi\kappa_0$
 is the sum of the partial density of states in
occupied subbands. Thus, the  Coulomb interaction in nanotubes is
screened rather weakly, $\overline{V}_0(z)\sim
e^2/(z\ln^2{(z/a)})$.

            To find  the oscillating part $\widetilde{V}_0(z)$ we
    consider a small vicinity of the point $k=2p_l$ and note that the factor $\tilde{e}^2I_0K_0$
    in the integrand of eq.~(\ref{V0(z,phi)}) cancels. Thus, we see yet another peculiarity
    of the 1D problem: Friedel oscillations of the screened potential do not
    depend on the charge! (see below eqs.~(\ref{tildeV}), (\ref{VNZ})). This is due to infinitely large magnitude of $\Pi_0(k)$ at $k=2p_l$.
    Then one can introduce the parameter
    $\xi=(k-2p_l)/2p_l$ and take the integral over $\xi$ from $-\infty$ till $\infty$.
     By the same shift of the integration path as described above we arrive
                  at the series in powers of $1/\ln{(4p_lz)}$:

\begin{equation}\label{tildeV}
    \widetilde{V}_0(z)=-\sum^{l=L}_{l=-L}\frac{2\pi^2p_l}{m}\frac{\cos(2p_lz)}
    {z\ln^2(4p_lz)}[1-\frac{2C}{\ln{(4p_lz)}}+ \ldots]
\end{equation}

We see that the amplitude of the Friedel oscillations in nanotubes
decreases with increasing distance not slower than the regular
part $\overline{V}_0(z)$ does. Moreover, the ratio
$\widetilde{V}_0/\overline{V}_0$ at any $z$ in the order of
magnitude can be estimated as $1/(p_Fa_B)$,  and for "metallic"
limit of the dense electron gas ($p_Fa_B\gg 1$) the Friedel
oscillations become negligibly small.

            Qualitatively different results are obtained for non-zero harmonics.
     The factor $I_n(ka)K_n(ka)$ for $n\neq 0$ tends to the constant $1/2n$
      for $k\rightarrow 0$
     and $\Pi(k,n)$ at $k=0$ remains finite either. Thus, all non-zero harmonics
     of the Coulomb interaction undergo the screening of the dielectric type,
     that is, their dependence on the distance $z$ coincides with that of the bare
      Coulomb law. Indeed, from eq.~(\ref{V0}) we have for the $n$-th harmonic of the bare
      interaction
\begin{eqnarray}\label{V_n}
    V_n^{(0)}(z)=\frac{\tilde{e}^2}{\pi a}Q_{n-1/2}(1+\frac{z^2}{2a^2});
    \\
 \nonumber   V_n^{(0)}(z\gg a)\simeq \frac{\Gamma(n+1/2)}{\sqrt{\pi}n!}
    \frac{\tilde{e}^2}{z}(\frac {a}{z})^{2n},
    \end{eqnarray}
where $Q_\nu$ is the spherical function of the second type. The
regular part of the screened  interaction stems again from the
region of small $k$. Eq.~(\ref{Pi}) gives:

\begin{eqnarray}\label{Pi_n}
    \mbox{Im}(\Pi(k,n)|_{k=0}=0,
 \nonumber  \\
    {\rm Re}(\Pi(k,n))=\frac{m}{2\pi^2
    k}\sum_{-L}^{L}\ln{\biggl|\frac{(k^2a^2+n^2+2kp_la^2)^2-4n^2l^2}{(k^2a^2+n^2-2kp_la^2)^2-
    4n^2l^2}\biggr|}
        \end{eqnarray}
and there are two possibilities for the quantity
$\kappa_n\equiv\Pi(0,n)$. If the subband occupation for $T=0$
terminates at $l=\pm L$ and $n^2>4L^2$, then

 \begin{equation}\label{kappa_n}
    \kappa_n=\frac{4ma^2}{\pi^2}\sum_{l=-L}^L \frac{p_l}{n^2-4l^2},
    \end{equation}

Otherwise, for  $n^2\leq 4L^2$  and $n$ even the $n$-th harmonic
of the potential couples the degenerate states $l_0=n/2$ and
$l_0=-n/2$. In this case we calculate the limit of the uncertainty
in eq.~(\ref{Pi_n}) at $n=2l_0$, $k\rightarrow 0$ and find
\begin{eqnarray}\label{l_0}
    \Pi_{n=2l_0}=\frac{2m}{\pi^2}(\frac{1}{p_{l_0}}+\frac{2p_{l_0}a^2}{n^2})
    \end{eqnarray}
After that the terms with  $l=\pm l_0$ in the sum of
eq.~(\ref{kappa_n}) should be replaced by $\Pi_{n=2l_0}$ from
eq.~(\ref{l_0}). Thus, for the $n$-th harmonic of the screened
potential at large $|z|$ we obtain the contribution from the point
$k=0$ :

\begin{equation}\label{V_n(z)}
    \overline{V}_n(z)= \frac{\Gamma(n+1/2)}{\sqrt{\pi}n!}
    \frac{\tilde{e}^2}{z}(\frac {a}{z})^{2n}(1+\frac{\pi\kappa_n}{ m a_Bn})^{-2}
    \end{equation}
The role of the  effective dielectric constant for the n-th
harmonic is played by the quantity
    \begin{equation}\label{epsilon}
    \epsilon_n=(1+\frac{\pi\kappa_n}{ ma_Bn})^2
    \end{equation}
that tends to 1 with increasing $n$. The dielectric type of
screening of harmonics with $n\geq 1$ can be understood in terms
of the classical electrostatics (suggested by M.V.Entin). Each
term of the series in eq.~(\ref{V(z,phi)}) corresponds to the one
of the expansion in multipoles: $n=1$  gives dipole-dipole
contribution, $n=2$ - quadrupole-quadrupole  and so on. At very
large $|z|$ the fields of all multipoles on the nanotube surface
are practically normal to the axis, hence, they cause only the
displacements of electrons in the azimuthal direction $\varphi$.
That is why the $z$-dependence  of the e-e  interaction does not
change; the system simply is polarized in accord with the
dielectric mechanism when only the bound charges are available.

        As to the oscillating part $\widetilde{V}_n(z)$
    it is determined by zeros of the argument of logarithm in the eq.~(\ref{Pi_n}).
     First of all one can see that such contributions exist not for all values
     of $n$. For a fixed  concentration of electrons $p_Fa\leq L$ whereas the
     singularity in the right-hand-side of eq.~(\ref{Pi_n}) occurs only if  $p_F^2a^2>(n-l)^2$
     with $|l|<L$. Thus, the  Friedel oscillations of the $n$-th harmonic exist for $n\leq 2L$
     but for $n>2L$ we have only non-oscillating  contribution $\overline{V}_n(z)$
given by eq.~(\ref{V_n(z)}). The calculation of the oscillating
part $\widetilde{V}_n(z)$ for $n\leq2L$ is totally similar to the
derivation of eq.~(\ref{tildeV}). The singular points $k_{c}$ are
given by the relation
 \begin{equation}\label{k_c}
    k_{c}=p_l\pm\sqrt{p_F^2-\left(\frac{n-l}{a}\right)^2}
    \end{equation}
and the polarization operator at $k\rightarrow k_{c}$ takes the
form
\begin{eqnarray}\label{Piq}
    \Pi\approx(\frac{m}{2\pi^2k_{c}})\ln{|\frac{q_l}{k-k_{c}}|};
    \\
 \nonumber   q_l=\frac{2k_{c}p_l(k_cp_la^2+nl)}{nl(k_{c}-p_l)}.
    \end{eqnarray}
The result reads
 \begin{equation}\label{VNZ}
    \widetilde{V}_n(z)=-\frac{2\pi^2}{mz}\sum_l\frac{k_c\cos(k_{c}z)}{\ln^2(zq_l)}
    \end{equation}
Thus, only in the case $n\leq 2L$ the Friedel oscillations in
nanotubes give a noticeable contribution at large distances that
is oscillating part of the  $n$-th harmonic of the screened
potential decreases slower with increasing $z$     as compared
with the regular part  $\overline{V}_n(z)$. It is worth to note
also that the Friedel oscillations look like superposition of the
monochromatic waves with various periods $2\pi/k_c$ (depending
both of  $n$  and $l$) rather than the single wave $\cos(2p_Fr)$
as it is in 2D and 3D systems.

        The problem of the screened Coulomb potential in nanotubes was considered in \cite{lin}.
However, the authors have given only numerical results for zeroth
harmonic $\overline{V}_0(z)$ and did not discuss the Friedel
oscillations.

        To conclude, we have analyzed the e-e  interaction   in nanotubes in the frames of the linear
     screening theory. Qualitative difference between  quasi-1D systems  and 3D and 2D systems
     is established. The  Friedel oscillations in nanotubes do not determine the asymptotic behaviour
     of the screened potential at large distances in contrast with 3D and 2D systems.

\acknowledgments This work has been supported   by the  RFBR, by
the RF President
     grant for scientific schools, as well as by the
 Programs of the Russian Academy of Sciences.


\begin{thebibliography}{5}
\bibitem{ando}
  \Name{Ando T., Fowler A. \and Stern F.}
  \REVIEW{Rev. Mod. Phys.}{54}{1982}{437}.

\bibitem{egger}
  \Name{Egger R., Grabert H.}
  \REVIEW{Phys.Rev.Lett.}{79}{1997}{3463}.

  \bibitem{yuan}
  \Name{Yuan Q., Chen H., Zhang Y. \and Chen Y.}
  \REVIEW{Phys.Rev.B}{58}{1998}{1084}.

  \bibitem{grab}
  \Name{Egger R., Grabert H.}
  \REVIEW{Phys.Rev.Lett.}{75}{1995}{3505}.

  \bibitem{schulz}
  \Name{Schulz}
  \REVIEW{Phys.Rev.Lett.}{71}{1993}{1864}.

\bibitem{gell}
  \Name{Gell-Mann M., Brueckner K.}
  \REVIEW{Phys.Rev.}{106}{1957}{364}.

\bibitem{lifshitz}
  \Editor{E.~M.~Lifshitz, L.~P.~Pitaevskii}
  \Book{Fizicheskaya kinetika}
  \Vol{10}
  \Publ{Nauka, Moskva}
  \Year{1979},
 \S{40}.

\bibitem{lin}
  \Name{Lin M.F., Chuu D.S.}
  \REVIEW{Phys.Rev.B}{56}{1997}{4996}.
\end{thebibliography}
\end{document}